\documentstyle[12pt]{article}
\def\boldvec#1{\mbox{\boldmath $#1$\unboldmath}}
\begin{document}
\centerline{{\Large\bf Dissipation and Topologically Massive }}

\vspace{.1in}
\centerline{{\Large\bf Gauge Theories in Pseudoeuclidean Plane}}

\vspace{.3in}
\centerline{\bf M.Blasone${}^{*}$, E.Graziano${}^{*}$, 
O.K.Pashaev${}^{\dagger}$
and G.Vitiello${}^{*}$}
 
\vspace{.2in}
\centerline{{\it ${}^{*}$Dipartimento di Fisica}}
\centerline{{\it  Universit\`a di Salerno, 84100 Salerno, Italy}}
\centerline{{\it and INFN Gruppo Collegato Salerno}}

\vspace{.2in}
\centerline{{\it  ${}^{\dagger}$Joint Institute for Nuclear Research}}
\centerline{{\it Dubna (Moscow) 141980, Russian Federation}}

\begin{abstract}
In the pseudo-euclidean metrics Chern-Simons gauge theory in the infrared 
region is found to be associated with dissipative dynamics.
In the infrared limit the Lagrangian of 2+1 dimensional 
pseudo-euclidean topologically massive electrodynamics has indeed
the same form of the Lagrangian of the damped 
harmonic oscillator. On the hyperbolic plane a set of two damped harmonic
oscillators, each 
other time-reversed, is shown to be equivalent to a single undamped 
harmonic oscillator. The
equations for the damped oscillators are proven to be the same as the 
ones for the Lorentz force acting on two particles carrying opposite charge 
in a constant magnetic field and in the electric harmonic 
potential. This provides an immediate link with 
Chern-Simons-like dynamics of Bloch electrons in solids
propagating along the lattice plane with hyperbolic energy 
surface. The symplectic structure of the reduced theory is finally 
discussed in the Dirac constrained canonical formalism.
\end{abstract}
 

\newpage

\noindent{\bf 1. Introduction}
\medskip

It is well known that quantum gauge field theory in 2+1 dimensional
space-time admits nontrivial topological structure
defined by the Chern-Simons term[1,2,3] and, in the euclidean metrics,
it provides
an useful theoretical framework in the description of
interesting phenomena
in planar condensed matter physics as, for example, fractional statistics 
particles[4], the quantum Hall effect, high $T_{c}$ superconductivity[5].

The topological properties of the theory are connected with the form
of Chern-Simons term, which is independent of the space-time metrics. The 
metric structure, however, is crucial in determinig other physical 
features which, as we will see, are related with the canonical structure of 
the theory.

In ref.[2] topological Chern-Simons theory has been studied by resorting 
to the quantum mechanical analysis of an oscillator model.
The purpose of this paper is to show that
in pseudo-euclidean space 
the Lagrangian density of 2+1 dimensional gauge theory with Chern-Simons 
term (topological mass term) and Proca mass term can be interpreted in the 
infrared region[6] as the Lagrangian density of the damped harmonic 
oscillator, provided the topological Chern-Simons mass parameter is 
identified with the damping parameter: In the pseudo-euclidean metrics the
Chern-Simons gauge theories can be therefore associated with dissipative 
dynamics.

Such a conclusion is interesting in several respects.

{}From 
one side indeed, dissipative systems are of great interest in high energy 
physics (in quark-gluons physics, for example), in the early universe 
physics as well as in many body theories, in phase transition phenomena, 
in quantum optics and in many practical applications of quantum field 
theory at non-zero temperature.
The possibility, then, to exhibit the gauge theory structure and 
topologically nontrivial features underlying
dissipative phenomena is very appealing. On the other side,  realizing 
vice versa that topologically massive electrodynamics in pseudo-euclidean 
metrics can be interpreted as a dissipative dynamics can be as well 
interesting since new perspectives open up for
Chern-Simons gauge theories with consequent much wider range of possible 
applications.

In recent years the problem of the canonical quantization of 
the damped harmonic oscillator in the operator formalism has been 
intensively
studied and some further light has been shed on quantum dissipative 
phenomena[7-9]. By resorting to the results of Bateman on the canonical 
formalism for partial differential equations with dissipative terms[10], of 
Umezawa, Takahashi and collaborators on finite temperature quantum field 
theory in the operator formalism[11] and of Feshbach and Tikochinski on the 
quantum mechanics of the damped harmonic oscillator[12], it has been shown 
that the canonical quantization
of the damped oscillator can be properly performed by doubling the
phase-space degrees of freedom in the framework of quantum field theory(QFT)
[7]. 
The space of the states has been shown to split into unitarily inequivalent 
representations of the canonical commutation relations, and
the non-unitary character of the irreversible time evolution is expressed 
as tunneling among such inequivalent representations.
The vacuum has been found to have the structure of  SU(1,1)
time-dependent coherent state and its 
statistical and thermodynamical properties have been recognized[7] to be 
the ones of the thermal vacuum in Thermo Field Dynamics[11].

The SU(1,1) group 
structure underlying the theory signals the r\^ole
played by the pseudo-euclidean metrics in the canonical formalism for 
quantum dissipation. In Sec.2 we indeed show the strict relationship 
between the doubling of the degrees of freedom for dissipative systems
and the pseudo-euclidean metrics.
In particular we prove that the system made of the damped harmonic 
oscillator and of its time-reversed image (the {\it doubled} damped 
oscillator) globally behaves as a closed system described by a single 
{\it undamped} harmonic oscillator. Vice versa, by use of hyperbolic 
coordinates (pseudo-euclidean metrics) an harmonic oscillator can be 
"decomposed" into two damped harmonic oscillators (each other
time-reversed). Moreover, we also show that by convenient representation  
of the dissipative factor by a magnetic field[13] and by choosing
pseudo-euclidean metrics the equations for the damped harmonic oscillator 
and of its time-reversed image can be cast in the form of the Lorentz force 
equations for particles of opposite electric charge $e_1 = - e_2 = e$, 
respectively.  In Sec.3 we show that, by considering its infrared 
properties, the Lagrangian for the topologically massive electrodynamics 
reproduces the Lagrangian for the damped harmonic oscillator with doubled 
degrees of freedom when the pseudo-euclidean metrics is used. As an 
explicit application we study in Sec.4 the dynamics of the Bloch electrons 
in solids propagating along the plane of the lattice with  hyperbolic energy
surface[14]. In this example the Lorentz force (the Chern-Simons term) 
clearly plays the r\^ole of the damping term and again the system 
Lagrangian density is recognized to be the one of the damped harmonic 
oscillator. In Sec.5 we analyze the canonical structure in the context of 
constrained canonical formalism of Dirac. Finally, Sec.6 is devoted to 
conclusive remarks.

\bigskip
 \bigskip
\noindent {\bf 2. The damped harmonic oscillator and the
pseudo-euclidean metrics}
\medskip

In order to introduce our subsequent discussion on the Chern-Simons term,
we need to show
how the pseudo-euclidean metrics arises in the canonical formalism for 
dissipative systems. We consider the simple case of one-dimensional 
damped harmonic oscillator (dho):
$$
m \ddot x + \gamma \dot x + k x  = 0  ~~, \eqno(2.1a)
$$
with time independent $m$, $\gamma$ and $k$.

It is well known[7-12] that to set up the canonical formalism for 
dissipative systems the doubling of the degrees of freedom is required in 
such a way
to complement the given dissipative system with its time-reversed image, 
thus obtaining a globally closed system for which the Lagrangian formalism 
is well defined. The time-reversed image of the given system plays the 
r\^ole of reservoir or thermal bath into which
the energy dissipated by the original system flows. In the case of 
system (2.1a) the doubling of the $x$ degree of freedom leads to consider
the dho in the {\it doubled} $y$ coordinate
$$
m \ddot y - \gamma \dot y + k y  = 0 ~~.\eqno(2.1b)
$$

The system of damped harmonic oscillators (2.1a) and its {\it 
time-reversed} $(\gamma \rightarrow - \gamma)$ image (2.1b) is then a closed 
system described by the Lagrangian density
$$
 L = m\dot x \dot y + {\gamma\over 2} (x \dot y -\dot x y) - k xy ~~.
\eqno(2.2)
$$
The canonically conjugate momenta $p_{x}$ and $p_{y}$ can now be 
introduced as customary in the Lagrangian formalism:
$$p_{x} \equiv {{\partial L}\over{\partial \dot x}} = m \dot y - 
\frac{\gamma}{2} y\;\;\;\;\;\;\;,\;\;\;\;\;\;
p_{y} \equiv {{\partial L}\over{\partial \dot y}} =
m \dot x + \frac{\gamma}{2}  x ~~, \eqno{(2.3)}
$$
and the dynamical variables $\{ x , p_{x} ; y , p_{y} \}$ span the new 
phase-space.
The Hamiltonian is then
$$
H = p_{x} \dot x + p_{y} \dot y - L = {{1}\over{m}} p_{x} p_{y} + 
{{1}\over{2m}}\gamma 
\left ( y p_{y} - x p_{x} \right ) + \left ( k -
{{\gamma^{2}}\over{4 m}} \right ) x y \quad . 
\eqno{(2.4)}
$$
The common frequency of the two oscillators (2.1a) and (2.1b) will be 
denoted by $\Omega \equiv \left [ {1\over{m}} \left ( k - 
{{\gamma^{2}}\over{4 m}} \right ) \right ]^{1\over{2}}$,
and is assumed to be
real, i.e. $k > {{\gamma^{2}}\over{4 m}}$~, corresponding to the case 
of {\it no} overdamping.

Canonical quantization may then be performed by introducing
the commutators 
$[\, x , p_{x} \, ] = i\, \hbar = [\, y , p_{y} \,] \quad , \quad  [\, x , 
y \,] = 0 = [\, p_{x} , p_{y} \, ]$, and the corresponding sets of 
annihilation and creation operators
$$a  \equiv \left ({1\over{2 \hbar \Omega}} \right )^{1\over{2}} \left ( 
{{p_{x}}\over{\sqrt{m}}} - i \sqrt{m} \Omega x \right ) \quad ; \quad 
a^{\dagger} \equiv \left ({1\over{2 \hbar \Omega}} \right )^{1\over{2}}
\left ( 
{{p_{x}}\over{\sqrt{m}}} + i \sqrt{m} \Omega x \right ) ;$$

$$b \equiv \left ({1\over{2 \hbar \Omega}} \right )^{1\over{2}} \left ( 
{{p_{y}}\over{\sqrt{m}}} - i \sqrt{m} \Omega y \right ) \quad ; \quad 
b^{\dagger} \equiv \left ({1\over{2 \hbar \Omega}} \right )^{1\over{2}}
\left ( {{p_{y}}\over{\sqrt{m}}} + i \sqrt{m} \Omega y \right ) , 
\eqno{(2.5)}
$$
with  
$$
[\, a , a^{\dagger} \,] = 1 = [\, b , b^{\dagger} \,] \quad , \quad  
[ \,a , b \,] = 0 = [\, a , b^{\dagger} \,] \quad . \eqno{(2.6)}
$$

A detailed account of the quantization procedure in QFT may be found in 
refs. [7,8].

We stress that time-reversal of $y$-oscillator with respect 
to $x$-oscillator implies that the $y$-oscillator acts indeed as the sink 
where the energy dissipated by the $x$-oscillator flows and thus as the 
reservoir or heat bath coupled to the $x$-oscillator\footnote{
The linear system with ''negative damping`` could be realized
physically, e.g., as the  Froude pendulum or the electronic feedback
generator
[25].}.
This means that in 
order to set up the canonical formalism for the damped oscillator (2.1a) 
the details of the reservoir dynamics need not to be specified; the only 
requirement to be met is that the reservoir must receive {\it all} the 
energy flux outgoing from the $x$-system so that the coupled
$(x-y)$-system  globally behaves as a closed (non dissipative) system: In 
other words, we can show that the set of eqs.(2.1) is formally equivalent 
to the equation for the
({\it undamped}) one dimensional harmonic oscillator, say $r(t)$, 
representing the global $(x-y)$-system:
$$
m \ddot r  + K r  = 0  ~~, \eqno(2.7)
$$
with
$$K \equiv m{\Omega}^{2} ~~. \eqno{(2.8)}$$

Such an equivalence is easily proven if one uses
hyperbolic coordinates ${{x_1}(t)}$ and ${{x_2}(t)}$:
$${{x_1}(t)} = r(t) cosh u(t) ~~, ~~ {{x_2}(t)} = r(t) sinh u(t)  ~~,
\eqno{(2.9)}
$$
in terms of which the oscillator coordinate $r(t)$ is indeed 
expressed in the pseudo-euclidean metrics:
$$r(t)^{2} \equiv {{x_1}(t)}^{2} - {{x_2}(t)}^{2} ~~. \eqno{(2.10)}
$$
Use of the (canonical) transformation
$$x(t) = {{{{x_1}(t)} + {{x_2}(t)}}\over{\sqrt 2}} = {1\over{\sqrt 2}}
r(t) e^{u(t)} ~~,
y(t) = {{{{x_1}(t)} - {{x_2}(t)}}\over{\sqrt 2}} =  {1\over{\sqrt 2}}
r(t) e^{-u(t)} ~~, \eqno{(2.11a)}
$$
in eqs.(2.1) readily gives eq.(2.7), provided 
$$u(t) \equiv -
{{\gamma}\over {2m}} t ~~, \eqno(2.11b)$$
as required by the time independence of the coefficients in eqs.(2.1).
Vice versa, the oscillator (2.7) is
{\it decomposed} into two damped oscillators (2.1) when the
pseudo-euclidean metrics (2.10) is adopted and the transformation (2.11a-b) 
is used.
Similarly, use of eqs.(2.11) in the Lagrangian (2.2) gives the Lagrangian 
for the harmonic oscillator (2.7):
$$
L={1\over 2}  m \dot r^2 - {1\over 2}K r^2
\eqno(2.12)
$$
(and vice versa (2.2) can be obtained from (2.12)).

Of course, if one chooses the euclidean metrics, $r^{2} \equiv x_{1}^{2} + 
x_{2}^{2} ~,~~ x_{1} = r cos \alpha ~,~~ x_{2} = r sin \alpha ~$, the
$r$-oscillator is {\it decomposed} into two {\it undamped} oscillators.

These observations, although of elementary simplicity, teach us the 
r\^ole of the pseudo-euclidean metrics in setting up the canonical 
formalism for dissipative systems and clarify the meaning of the doubling 
of the degrees of freedom.

In the presence of the damping factor $e^{-\Gamma t}$ forward time 
evolution cannot be mapped (transformed) into backward time evolution 
by any operation (as complex conjugation in the non-dissipative case) 
except time-reversal $t \rightarrow -t$ (or, which is the same, by changing 
the relative sign of $\Gamma$ and $t$; here we keep fixed the sign of 
$\Gamma$). In dissipative systems, however,
time-reversal symmetry is broken, which means that one cannot use
time-reversal $t \rightarrow -t$ without changing the physics; the conclusion
is that dissipation induces a {\it partition} 
on the time axis thus implying that positive 
and negative time directions necessarily must be associated with {\it 
separate} modes, respectively describing different physical situations. 

The canonical formalism, which only describes closed systems, needs
therefore to be constructed on a set of {\it double} modes, one for each 
direction of time evolution
(each other time-reversed), and these modes are indeed 
provided by hyperbolic coordinates (see eqs.(2.11)).

In the non-dissipative case, where only oscillating factors of type 
$e^{\pm iEt}$ 
are involved, one can actually limit himself to consider, e.g., only 
forward time direction, the backward time direction being obtained by 
complex conjugation operation (or by $t \rightarrow -t$ which is now 
allowed since
time-reversal is not broken); this is why one does not need to consider 
backward and forward modes as {\it separate } modes in the non-dissipative
systems.

We can conclude that the {\it decomposition} of a canonical (Hamiltonian) 
system into two subsystems needs pseudo-euclidean metrics when these 
subsystems are noncanonical ones. Vice versa, setting up the canonical 
formalism for a dissipative system requires (as well known) the inclusion 
into the formalism of the reservoir and this can be achieved by doubling 
the system (the degrees of freedom) which in turn implies the 
pseudo-eulidean metrics.
We also remark that the {\it decomposition} operation can be shown to be 
formally expressed at the algebraic level by the co-product operation of 
Hopf algebras[15]. We will not discuss this point which is out of the task 
of the present paper. Details on the relation with coalgebra and quantum 
deformation of Lie groups may be found in refs.[16,17].
It should be finally stressed that doubling of degrees of freedom is
a central ingredient not only 
in dissipation theory but also in thermal field theories[11,18] and in the 
$C^*$-algebraic formalism for statistical mechanics[19].

In terms of the hyperbolic coordinates $x_1$ and $x_2$ the Lagrangian (2.2) 
is
$$
L= L_{0,1} - L_{0,2}  + {\gamma\over 2}({\dot x_1} x_2 - {\dot x_2} 
x_1) ~~, \eqno(2.13)
$$
with
$$L_{0,i} =  {m\over 2} \dot x_i^2 - {k\over 2} x_i^2 ~~, i=1,2.
\eqno(2.14)
$$
The associate momenta are
$$
p_1 = m \dot x_1 + {\gamma\over 2} x_2 ~~,~~
p_2 = -m\dot x_2 -{\gamma\over 2} x_1 ~~ \eqno(2.15)
$$
and the motion equations corresponding to the set (2.1) are
$$
m \ddot x_1 + \gamma \dot x_2 + k x_1  = 0  ~~, \eqno(2.16a)
$$
$$
m \ddot x_2 + \gamma \dot x_1 + k x_2  = 0  ~~. \eqno(2.16b)
$$
The Hamiltonian (2.4) becomes
$$
H = H_1 - H_2 = {1 \over 2m} (p_1 - {\gamma\over 2}x_2)^2
+ {k\over 2} x_1^2 - {1 \over 2m} (p_2 + {\gamma\over 2}x_1)^2 -
{k\over 2} x_2^2 ~~. \eqno(2.17)
$$
It is interesting to write down (2.17) for the generic metrics
$g^{ij}= g_{ij}= diag(1, \kappa ^{2})$,
where $\kappa ^{2}= +1$ for euclidean spatial plane, and $\kappa ^{2}= - 1$ 
for pseudo-euclidean plane:
$$
H =  {1 \over 2m} (p_1 + {\kappa}^{2}{\gamma\over 2}x_2)^2
+ {k\over 2} x_1^2 + {1 \over 2m}{\kappa}^{2} (p_2 -
{\kappa}^{2} {\gamma\over 2}x_1)^2 +
{\kappa}^{2}{k\over 2} x_2^2 ~~. \eqno(2.18)
$$

Eq.(2.13) shows that the dissipative term actually acts as a coupling 
between the oscillators $x_1$ and $x_2$ on the hyperbolic plane. On the 
other hand, eq.(2.17) shows that the damping manifests as a correction in 
the kinetic energy for both oscillators.
In refs.[7,12] the Hamiltonian of quantum dissipation is obtained from 
eq.(2.4) (or (2.17)) by using the canonical operators (2.5) and it is
shown[7] to be the same as the one of Thermo Field Dynamics[11] for the 
boson case. In both cases, in dissipative dynamics as well as in Thermo 
Field Dynamics, the underlying group structure is the one of SU(1,1), the 
free Hamiltonian being the SU(1,1) Casimir operator[7]:
$H_{0,1} - H_{0,2}$.

In eqs.(2.13) and (2.17) it is remarkable the occurrence of
the minus sign between $L_{0,1}$ 
and $L_{0,2}$ and
$H_{1}$ and $H_{2}$ , respectively, which is in fact derived as a direct 
consequence of the
pseudo-euclidean metrics (cf.(2.18))
(such a minus sign is, on the contrary, postulated in Thermo Field 
Dynamics). In quantum dissipation and in
Thermo Field Dynamics the degeneracy of the ground state is strictly 
related to this special form of the free Hamiltonian[7,11]. We observe that 
no problem arises with the boundness from below of H in (2.17) since it is 
constant of motion and once its positive value is given it is preserved in 
time.

We note that the generator of the canonical transformations (2.9) is[20]
$${\cal U}(p_1 , p_2 ; r, u) \equiv p_1 r cosh u  +  p_2 r sinh u ~~,
\eqno(2.19)
$$
such that $x_{i} = {{\partial {\cal U}}\over{\partial p_{i}}}$ , $i=1,2$, 
and $p_r = {{\partial {\cal U}}\over{\partial r}} = p_1 cosh u +
p_2 sinh u $ , $p_u = {{\partial {\cal U}}\over{\partial u}} = p_1 r sinh u 
+ p_2 r cosh u $. One then easily derives the invariant
form
$${p_1}^{2} - {p_2}^{2} = {p_r}^{2} - {1\over{r^2}} {p_u}^{2} ~~, \eqno(2.20)$$
where again we notice the appearance of the minus sign. By observing that 
$p_r = m \dot r$ and $p_u = 0$ and by adding to both members of (2.20) the 
invariant (under (2.9)) quantity
${k\over2} r^{2}$ we recover again (2.17) and its equality with the 
Hamiltonian of the harmonic oscillator $r$:
${1\over 2}  m \dot r^2 + {1\over 2}K r^2$ .

By following ref.[13] we now introduce the vector potential
$$
A_i =  {B\over 2} \epsilon_{ij} x_j ~~,~~ i,j = 1,2~,~~B \equiv 
{c\over{e}}\gamma ~~,\eqno(2.21)
$$
with ${\epsilon}_{ii} = 0~, {\epsilon}_{12} = - {\epsilon}_{21} = 1$~, and 
the magnetic field
$\boldvec{ B} = \boldvec{\nabla} \times 
\boldvec{A} = - B \boldvec{{\hat 3}}$~.
Then $H_i$, $i=1,2$,
in eq.(2.17) describe two particles with opposite charge
$e_1 = - e_2 = e$
in the (oscillator) potential
$\Phi \equiv {k\over 2e}({x_1}^2 - {x_2}^2) \equiv {\Phi}_1 - {\Phi}_2 ~, 
{\Phi}_i \equiv {k\over2e}{x_i}^{2}$
and in the constant magnetic field $\boldvec{B}$ ~:
$$
H = {H_1} - {H_2} = {1 \over 2m} (p_1 - {e_1 \over{c}}{A_1})^2  + 
{e_1}{\Phi}_1
- {1 \over 2m} (p_2 + {e_2 \over{c}} A_2)^2 + {e_2}{\Phi}_2 ~~.\eqno(2.22)
$$
In fact eqs.(2.16) are nothing but the expressions of the Lorentz forces on 
particles with charge $e_1 = - e_2 = e$, in the electric field
$\boldvec{ E} = - \boldvec{\nabla } \Phi $~,  
and in the magnetic field $\boldvec{B} = - B  \boldvec{{\hat 3}}$:
$$
{\cal F}_{i} = m\ddot x_i = {e_i}[E_i + {1\over{c}}(\boldvec{v} \times 
\boldvec{B})_i] ~~,~ i =1,2~, \eqno(2.23)$$
with $\boldvec{v} = (\dot x_1 , \dot x_2 , 0)$. By using (2.21) the
Lagrangian (2.13) is rewritten in the familiar form
$$
L =  {1 \over 2m} (m{\dot x_1} + {e_1 \over{c}} A_1)^2
- {1 \over 2m} (m{\dot x_2} + {e_2 \over{c}} A_2)^2
- {e^2\over 2mc^2}({A_1}^2 + {A_2}^2) - e\Phi
$$
$$
= {m \over 2} ({\dot x_1}^2 - {\dot x_2}^2) +{e\over{c}}( {\dot x}_1 A_1 + 
{\dot x}_2 A_2) - e{\Phi} ~~, \eqno(2.24)
$$
from which eqs.(2.23) are
derived in the form
$$
{d\over {dt}}(m\dot x_i + {{e_i}\over c} A_i) = - {e_i}{\partial 
_i}{\Phi}_{i}
 + {{e_i}\over c} {\partial _i}{v_j} A_j ~~,~i\neq j ~, no~~ sum~~ on~~ 
 i,j~,\eqno(2.25)
 $$
where ${\partial _i} $ denotes ${\partial \over {\partial x_i}}$. Note
that ${d\over{dt}}A_i = v_j{\partial _j}A_i$~.

In the following sections we consider the relation with Chern-Simons 
gauge term and we study the dynamics of Bloch electrons in solids as a
physical application.

\bigskip
\bigskip

\noindent {\bf 3.   Topologically   massive    Chern-Simons    gauge    
theory in pseudo-euclidean space}
\medskip

We consider 2+1 dimensional space-time with a metric tensor
$$
g^{\mu \nu }= g_{\mu \nu }= diag(1, - 1, - \kappa ^{2}) ~~,
\eqno(3.1)$$
where $\kappa ^{2}= +1$ for euclidean space  plane,  and 
$\kappa ^{2}= - 1$  for  pseudo-
euclidean plane. Then the Abelian gauge field theory is  described  by
the Maxwell form
$$
L_{M} = - {1\over 4} F_{\mu \nu }F^{\mu \nu } = {1\over 2} \{(F^{2}_{01} + 
\kappa ^{2}F^{2}_{02} ) - \kappa ^{2}F^{2}_{12} \} ~~,
\eqno(3.2)$$
where
$$
F _{\mu \nu } = \partial _{\mu }A_{\nu } - \partial _{\nu }A_{\mu } ~~.
\eqno(3.3)$$

For the description of the infrared limit[6]
usually a gauge non-invariant
mass term can be introduced in the Proca form[21]
$$
L_{P} = {1\over 2} \mu ^{2} A_{\mu } A^{\mu } = 
{1\over 2} \mu ^{2}(A^{2}_{0} - A^{2}_{1} - \kappa ^{2} A^{2}_{2} ) ~~.
\eqno(3.4)$$
On the other hand, the Chern-Simons term
with statistical parameter $\theta$  acts as a  gauge  invariant 
topological  mass  term[1-5]
$$
L_{CS} = {\theta\over 2} \epsilon ^{\mu \nu \lambda } 
A_{\mu } \partial _{\nu }A_{\lambda } ~~.
\eqno(3.5)$$
We note that the Chern-Simons term (3.5) does not depend on the
(euclidean or pseudo-euclidean) metrics
 one chooses.

General, three-dimensional topological massive
Maxwell-Chern-Simons-Proca
electrodynamics in the Weyl gauge, $A_{0} = 0 $,
has the Lagrangian density
$$
L = L_{M} +  L_{CS} + L_{P} = {1\over 2} \{(\dot{A}^{2}_{1} + 
\kappa ^{2}\dot{A}^{2}_{2}) - \kappa ^{2}(\boldvec{\nabla} \times 
\boldvec{A})^{2}\} $$
$$ +{\theta\over 2} |\boldvec{ \dot{A}} 
\times \boldvec{ A}| - {1\over 2}  \mu 
^{2}(A^{2}_{1} + \kappa ^{2} A^{2}_{2} )  ~~.\eqno(3.6)$$

We now remark that, despite its independence of the metrics, 
the {\it topological}  term leads  to
drastically different situations depending if one works in the 
euclidean , $\kappa ^{2}= +1,$ or in pseudo-euclidean , $\kappa ^{2}= - 1,$ 
space metrics.

In the euclidean case  we  have  topologically  massive
electrodynamics[1-5]. For very massive modes
when $\theta  \rightarrow \infty $ the leading contribution is from the 
Chern-Simons term. It is well known that (euclidean) Chern-Simons gauge 
theory  with matter field describes particles
with  fractional  statistics - anyons[1-5].
In this case the topological  mass
parameter $\theta $ plays the r\^ole of  statistical  parameter  and
defines the spin of the particles.

Our statement is now that in the pseudo-euclidean case the 
Chern-Simons term generates dissipation
in the infrared limit. Indeed, infrared  properties  of
the theory are related to the constant in space
vector potential $q_{i}(t)$:
$$
A_{i}(x,t) = q_{i}(t) + \tilde{A}_{i}(x,t)~~,~~for~ \boldvec{ x} \rightarrow 
{\infty}~~ with~{\tilde A}_{i}(x,t) \rightarrow 0 ~~,
\eqno(3.7)$$
and the corresponding  Lagrangian  density  (3.6)
becomes in the infrared limit
$$
L= {1\over 2} (\dot{q}^{2}_{1} + \kappa ^{2}\dot{q}^{2}_{2}) + 
{\theta\over 2} |\boldvec{ \dot{q}} \times \boldvec{ q}| - {1\over 2} \mu 
^{2}(q^{2}_{1} + \kappa ^{2} q^{2}_{2} ) ~~.~
\eqno(3.8)$$

For $\kappa ^{2}= - 1$ this $L$  has exactly  the  form  (2.13) of the 
Lagrangian density for the damped
harmonic oscillator, provided the  Proca  mass is identified with the 
oscillating  frequency: $ {\mu}^2 \equiv \omega ^{2}= k/m$  , and the  
topological
(Chern-Simons) mass $\theta $ is related to the damping parameter $\gamma$:
$$
\theta  = {\gamma\over m} ~~. \eqno(3.9)$$

We thus conclude that damping has a global,  topological
origin and cannot be obtained by usual local mass.

In  this  context
interesting problems arise if we consider Chern-Simons  gauge field 
interacting
with matter fields in the pseudo-euclidean plane. The
dynamics  of  charged  particle  moving  in the hyperbolic plane is
studied in connection with problems of chaotic dynamics[22,23].
A connection exists between the classical behaviour of a 
system  and  its
energy level fluctuations. Typical problems are the ones of the billiard
in magnetic field with time-reversal symmetry breaking[23]. In the presence 
of time-reversal breaking and
non-trivial flux (Aharonov-Bohm  potential) the spectrum is entirely 
controlled by topology.  When  the
classical trajectories are  chaotic, such time-reversal
symmetry  breaking  has  a
dramatic effect on the spectrum: it changes the universality class  of
the local statistics of high-lying energy levels[23]. Switching on
of the flux leads to switching off the time-reversal invariance  without  
changing
the geometry of classical trajectories. If the chaotic motion  of  the
system, like $K$-systems[24], with exponential  separation  of  trajectories
with time can  be simulated by motion in the hyperbolic  plane,  and
the  Aharonov-Bohm statistical flux can be introduced by Chern-Simons
term, the  problem is formally very  close  to the framework above 
discussed. 
Indeed, the hyperbolic plane, as a simplest surface of constant negative
curvature, can be globally embedded in a space endowed with a Minkowskian
metrics instead of a Euclidean one (for surfaces of
negative curvature such a global embedding in Euclidean space is in fact
prohibited [26]. On the other hand only negative
curvature leads to chaotic motion[27]).
To be more specific on this point we study in the following 
section the case of hyperbolic  energy  surface in the dynamics of 
electrons in solids propagating along the plane of the lattice.

\bigskip
 \bigskip
\noindent {\bf 4. Dynamics of Bloch Electrons}
\medskip

As an application of physical interest we consider in this section
the  dynamics  of Bloch electrons  in  solid. As well 
known[14], in such a problem a central r\^ole is played by the inverse 
effective mass tensor
$$
{1\over m} \rightarrow {1\over {\hbar ^{2}}}
{\partial ^{2}\over {\partial \boldvec{ k}}^{2}}E(\boldvec{ k})~~.
\eqno(4.1)$$

In semiconductors the energy surface $E(\boldvec{ k})$ has the form
$$
E(\boldvec{ k}) = \hbar ^{2}( {k^2_1\over 2m_1} + 
{k^{2}_{2}\over 2m_{2}} + {k^{2}_{3}\over 2m_{3} })
\eqno(4.2)$$
in a convenient coordinate system. The tensor (4.1) is not necessarily  
positive
definite. When the Fermi surface is near the band boundary, the sign of the 
mass $m_{3}$ is negative[14]. In this {\it hyperbolic} case, due
to  strong  Bragg  reflection  from  the  boundary  of  the   band,
the  electron  propagates along trajectories which are parallel
to  the  planes  of  the lattice[14] and
$E(\boldvec{ k})$ has "saddle  points", where, i.e., the curvature of the 
surface may be positive in one direction and negative in another one.
In Ziman's words[14], "the dynamics of states in this  region  may  be
most unexpected". Indeed, as we show below the dynamics in  this  case
has "dissipative" character.

We consider only 2-dimensional motion, which is the case when one of the  
mass,  for
example $m_{2}$, goes to infinity. We also restrict ourselves 
with the simple  case $m_{1} =
|m_{3}|$. Thus, redefining $m_{3} \rightarrow m_{2}$ , 
we have the energy surface
$$
E(\boldvec{ k}) = \hbar ^{2}( {k^{2}_{1}\over 2m} + \kappa ^{2}
{k^{2}_{2}\over 2m} )~~,
\eqno(4.3)$$
and the effective mass matrix
$$
M^{-1} = diag({1\over m}, {\kappa ^{2}\over m})~~.
\eqno(4.4)$$
\par
Since  Bloch  electrons are subject to the Lorentz  force[14], we have the 
Lagrangian for a particle in external  magnetic field
$B_3 =  \epsilon_{3ij}{\partial}_{i}A_j$~ and  electric field
$- \partial _{i}\Phi$~:
$$
L= {m\over 2}(\dot{x}^{2}_{1} + \kappa ^{2}\dot{x}^{2}_{2}) + 
{e\over{c}}({\dot x}_1{A}_1(\boldvec{ x}) + {\dot x}_2{A}_2(\boldvec{ x})) - 
e\Phi(\boldvec{ x})
~~.~\eqno(4.5)  ~~$$
 For rotationally symmetric motion in 2 dimensions $A_{i}(\boldvec{ x})$
and $\Phi(\boldvec{ x})$ are given by
$$
A_{i}(\boldvec{ x}) = \epsilon _{ij}x_{j} A(x)~~, \eqno(4.6)
$$
$$
\Phi(\boldvec{ x}) = \Phi(x)~~.\eqno(4.7)
$$

For simplicity we consider a constant magnetic field $A(x)= {B\over 2}$
and a quadratic scalar potential
$$
\Phi(x) = {k\over 2e} (x^{2}_{1} + \kappa ^{2} x^{2}_{2} )~~.
\eqno(4.8)$$
Then from (4.5)
$$
L= {m\over 2}(\dot{x}^{2}_{1} + \kappa ^{2}\dot{x}^{2}_{2}) + {e\over 2c}B 
|{\boldvec{\dot x}} \times 
\boldvec{ x}| - {k\over 2} (x^{2}_{1} + \kappa ^{2} 
 x^{2}_{2} )
\eqno(4.9)$$
and we have the same particle problem as in Sec.3, eq.(3.8). For $\kappa 
^{2}= -1$ it
is defined by the Lagrangian (2.13) (or equivalently (2.24)) and the 
magnetic field $B$ plays role of the damping $\gamma $~($B \equiv 
{c\over{e}} \gamma $).

\bigskip
 \bigskip
\noindent {\bf 5. Canonical structure and Chern-Simons limit}
\medskip

In previous sections we have obtained the Lagrangian (2.13) (or (2.24)) 
from the Chern-Simons
gauge theory in the infrared limit (Sec.3)
and from the Bloch electron dynamics (Sec.4). Therefore
we can restrict to (2.13) the discussion of the canonical structure.

The canonical momenta are defined by eqs.(2.15) and the Hamiltonian
is given by (2.17), while the canonical brackets are
$$
\{x_{i}, p_{j}\} = \delta_{ij} ~~,\eqno(5.1) $$
$$
\{x_{i}, x_{j}\} = \{p_{i}, p_{j}\} = 0 ~~.\eqno(5.2)
$$

To extract the effects of the pure damping we consider  the limit 
of strong damping, $\gamma >> m$~. In this case we can neglect the kinetic 
term in the
Lagrangian (2.13) and we have
$$
L = + {\gamma\over 2} ({\dot x_{1}}  x_{2} - x_1 {\dot  x_{2}})
- {k\over 2} (x^{2}_{1} - x^{2}_{2}) ~~.~ \eqno(5.3)
$$

The Euler-Lagrange equations now are
$$
\gamma {\dot x_{1}} + k x_{2} = 0 ~~,
$$
$$
\gamma {\dot x_{2}} + k x_{1} = 0  ~~,\eqno(5.4)
$$
with general solution
$$
x_{1}(t) = a e^{\lambda t} + b e^{-\lambda t} ~~, $$
$$
x_{2}(t) = -a e^{\lambda t} + b e^{-\lambda t} ~~,\eqno(5.5)
$$
where $ \lambda = {k\over \gamma} $. This solution describes the damping
of the coordinate $x = {1\over {\sqrt 2}}(x_{1} + x_{2})$ and the growing 
of the coordinate $y = {1\over {\sqrt 2}}(x_{1} - x_{2})$~.
It is worth to note that the damping coefficient
$\lambda = {k\over \gamma}$ in the limiting case is different from the
original value $\Gamma = {\gamma\over 2m}$.

Another way to proceed simply consists in putting the mass $m = 0$ in eq.(2.13)
(and (2.16)).
We need however to comment upon some peculiarity of this
reduction procedure [25]. Under the reduction,
4-dimensional phase space for the
second-order system (2.16) turns into 2-dimensional one $\{x_{1}, x_{2}\}$
for the
first-order system (5.4). The first-order
system is called the ''degenerate system``[25]: An arbitrary initial value
problem for (2.16), in general, does not apply to the degenerate system (5.4).
For eqs.(2.16), at $t = 0$ we can attach an arbitrary value
for coordinates $x_{1}, x_{2}$ and related velocities  $\dot x_{1},
\dot x_{2}$. However, only after some time interval we can describe the same
physical system by eqs.(5.4) and, moreover, in such a case $\dot x_{1}$ and
$\dot x_{2}$ cannot be arbitrary since they are completely determined by
given coordinates $x_{1}, x_{2}$, according to (5.4)(see eq.(5.8)).
Subsequent evolution
of the system is determined by the potential (5.6) only, and depends
on the signature of the corresponding bilinear form. When the mass m goes to
zero,
transition from a state incompatible with eq.(5.4) to a compatible one
is very fast. Acceleration at the initial time is very high (related
velocity is changing very fast). The transition to the massless limit
can be well approximated by the discontinuous jumping condition [25]:
the energy of the system cannot be changed by jump. In our case, all energy
is concentrated in the potential (5.6) (the kinetic
energy is vanishing in the massless limit $m \rightarrow 0$). The jumping 
condition thus implies
that under the jumping the coordinates of the system remain invariant and
only the velocities  can be changed.

The limiting case has analogs in the theories described above in
Secs.3 and 4. In terms of the topologically massive 
Maxwell-Chern-Simons-Proca  
theory (3.6) it corresponds
to the so called Chern-Simons limit, valid for long wave excitations.
Since pure Chern-Simons theory is meaningful only for topologicaly
nontrivial 3-manifolds and the Chern-Simons term
does not contribute to the 
energy-momentum tensor and has vanishing Hamiltonian, to have local,
propagating modes we must keep the Proca mass term (3.4).

For the Bloch electron model (4.5) the strong damping limit corresponds to 
strong external magnetic and electric fields, when we can neglect
the electron effective mass m. 

Since the reduction procedure for Chern-Simons theories
has already been discussed
in the literature[2], we only discuss the reduction 
in the context of the damping problem.

The reduced theory is described by the first-order Lagrangian (5.3).
The Hamiltonian is simply
$$
H = {k\over 2} (x^{2}_{1} - x^{2}_{2} )~~ \eqno(5.6)
$$
and, as shown below, the symplectic structure of the
reduced model (5.3) is
$$
\{x_{i}, x_{j}\} = {1\over \gamma} \epsilon_{ij} ~~. \eqno(5.7)
$$
The Hamilton equations of motion are
$$
{\dot x_{1}} = \{H, x_{1}\}
= - {k\over \gamma} x_{2} ~~,~~ {\dot x_{2}} = \{H, x_{2}\} = - {k\over 
\gamma} x_{1} ~~,
\eqno(5.8)$$
coinciding with the Euler-Lagrange equations (5.4).

The bracket (5.7) follows from the canonical one (5.1) by Dirac
procedure for constrained systems. Indeed, 
in the limit $m \rightarrow 0$ from eqs.(2.15) for
the canonical momenta we have the two following constraints
$$
C_{1} = p_{1} - {\gamma\over 2} x_{2} ~~, ~~ C_{2} = p_{2} + {\gamma\over 
2} x_{1} ~~,
\eqno(5.9)
$$ 
which are  second class as
$$
G_{ij} = \{C_{i}, C_{j}\} = - \gamma \epsilon_{ij} ~~.
\eqno(5.10)$$

Using the definition of Dirac brackets :
$$
\{A,B\}_{D} = \{A,B\}_{P} - \{A,C_{i}\}(G^{-1})_{ij}\{C_{j},B\} ~~,
\eqno(5.11)
$$
where
$$
(G^{-1})_{ij} = {1\over \gamma} \epsilon_{ij} ~~, \eqno(5.12)
$$
we obtain the following Dirac brackets
$$ \{x_{i},x_{j}\}_{D} = {1\over \gamma} \epsilon_{ij} ~~, \eqno(5.13a)
$$
$$
\{p_{i},p_{j}\}_{D} = {\gamma\over 4} \epsilon_{ij} ~~, \eqno(5.13b)
$$
$$
\{x_{i},p_{j}\}_{D} = {1\over 2} \delta_{ij} ~~. \eqno(5.13c)
$$

{}From the first brackets we recognize eq.(5.7) as the Dirac
bracket for constrained damped oscillator model (2.13).

Let us conclude this section with few remarks.

We observe that the pseudo-euclidean character of the kinetic term
in the original model (2.13) does not affect the constrained theory. 
The essential difference
between the euclidean case and the pseudo-euclidean one
relies in the sign of the second term in the Hamiltonian (5.6), i.e. in 
the
signature of the quadratic form.
The bracket (5.7) is the same in both the metrics
and is independent of the structure (5.6). However, the physics is 
drastically different in the two cases: in the euclidean metrics we have 
oscillating motion,
while in the pseudo-euclidean metrics we have damping phenomena.
The global, topological nature of the damping phenomena is shown by 
the non-analytic dependence on the damping $\gamma$ of the bracket (5.7) (or 
of the Hamilton equations (5.8)).

The above remarks suggest few more observations  on Chern-Simons
gauge field
theory and on the electron motion in external field.

For the gauge theory in the pure Chern-Simons limit, an addition
of anisotropic Proca 
mass term can lead to unusual phenomena for the light behaviour.

For the electron motion in a strong magnetic and electric
fields, the dynamics is independent of the electron effective mass
and is completely determined by the structure of the electric field. For 
the harmonic electric potential with positive signature
(the euclidean case),
the dynamics has an oscillating
character. While for a "saddle point" potential 
the harmonic force  with 
opposite coupling in different directions
leads to damping mode in one direction
and growing mode in the another one. Actually, the solution
(5.5) describes the motion on the hyperboloid
$$
x^{2}_{1}(t) - x^{2}_{2}(t) = 4ab = const ~~. \eqno(5.14)
$$

\bigskip
 \bigskip

\noindent {\bf 6. Conclusions}
\medskip

In this paper we have shown that
in the pseudo-euclidean metrics the Lagrangian density of 2+1 
dimensional gauge theory with
Chern-Simons term and Proca mass term has, in the infrared region, the same 
form of the Lagrangian density of the damped harmonic oscillator.
In the pseudo-euclidean metrics Chern-Simons gauge theory in the infrared 
region can be thus associated with dissipative dynamics. As an explicit 
example of practical interest, we have considered the propagation of
Bloch electrons in solids along the plane of the lattice with hyperbolic 
energy surface.

In order to exhibit the r\^ole played by pseudo-euclidean metrics in
the canonical formalism for
dissipation we have shown that the system of a damped harmonic oscillator
and of its time-reversed image, this last one introduced by doubling the 
degrees of freedom as required indeed by canonical formalism, actually 
behaves as a closed system described by a single harmonic oscillator. More 
generally, it is also true that by use of hyperbolic coordinates a closed 
system may be
"decomposed" into two open subsystems, each one acting
as the reservoir for the other one.

The equations for the damped 
oscillator and for its time-reversed image have been proven to be the same 
as the ones for the Lorentz force acting on two particles carrying opposite 
charge $e_1 = - e_2 = e$ in a constant magnetic field 
(simulating
the damping factor) and in the electric harmonic potential. Such a 
representation provides an immediate link with the above mentioned 
example of Bloch electrons in solids and with 
the dynamics of charged particle in the hyperbolic plane, which is 
studied also in connection with problems of chaotic motion[23].

Our analysis has been carried on at the classical level. However, the
canonical quantization of the damped harmonic oscillator has been worked 
out in the operator formalism in ref.[7] and its relation with
Feynmann-Vernon functional integral scheme and with Schwinger formalism for 
dissipative systems has been analysed in ref.[8].
As shown in [7] the set of states of the system splits into unitarily 
inequivalent representations of the canonical commutation relations and 
the irreversibility of time evolution is expressed in terms of tunneling 
among the unitarily inequivalent representations. The underlying group 
structure of the theory is the one of $SU(1,1)$ and the vacuum state is 
an $SU(1,1)$ time dependent coherent state.
In view of the results presented in this paper we expect that the canonical 
quantization scheme of refs.[7-9] can be extended to
the Maxwell-Chern-Simons-Proca gauge theory in the infrared region in 
pseudo-euclidean metrics.
We thus expect that also in the case of
Chern-Simons theories in pseudo-euclidean metrics a proper quantization
scheme requires the full set of unitarily inequivalent representations of 
the canonical commutation rules. In the same way, the thermal nature of 
the vacuum state in the damped oscillator case[7] and its relation with 
thermal vacuum state of quantum field theory at finite temperature in
real time (Thermo Field Dynamics[11]) may lead to interesting features in
extending the canonical quantization to pseudo-euclidean
Chern-Simons gauge theories.

Finally, we observe that the time evolution generator of
damped oscillator can be expressed in terms of quantum deformation of 
the
Weyl-Heisenberg algebra[16] and that the coproduct operation of
quantum deformed Lie-Hopf algebras play a relevant r\^ole in thermal field 
theories[17]. From the discussion of this paper it is then to be expected 
that quantum deformations of Lie-Hopf algebras also play a relevant 
r\^ole in pseudo-euclidean Chern-Simons theories. On this subject we plan 
to present our analysis in a future work.

This work has been partially supported by EU Contract ERB CHRX CT940423.

\newpage

\noindent {\bf References}
\bigskip
\begin{itemize}
\item[1]  S.Deser, R.Jackiw and S.Templeton, {\it Annals of Phys. (N.Y.)} 
{\bf 140} (1982) 372

\item[2] G.V.Dunne, R.Jackiw and C.A.Trugenberger, {\it Phys. Rev.} {\bf 
D41} (1990) 661 

\item[3] E.B.Park, Y.W.Kim, Y.J.Park, Y.Kim and W.T.Kim, {\it Mod. Phys. 
Lett.} {\bf  A10} (1995) 1119

\item[4]  F.Wilczek, {\it Phys. Rev.Lett.} {\bf 48} (1982) 1144;
  {\bf 49} (1982) 957 \\
F.Wilczek, {\it Fractional statistics and anyon
superconductivity}, World Sci., Singapore 1990\\
A.Lerda, {\it Anyons: Quantum mechanics of particles 
   with fractional statistics}, Springer-Verlag 1992

\item[5]  I.Dzyaloshinskii, A.M.Polyakov and P.B.Wiegmann, {\it Phys. 
Lett.} {\bf A127} (1988) 112 \\
P.B.Wiegmann, {\it Phys. Rev. Lett.} {\bf 60} (1988) 821\\
A.M.Polyakov {\it Mod. Phys. Lett.} {\bf A3} (1988) 325 \\
S.Bahcall and L.Susskind, {\it Int. J. Mod. Phys.} {\bf 
B5} (1991) 2735

\item[6]A.J.Niemi and V.V.Sreedhar, {\it Phys. Lett.} {\bf B336} (1994) 
381

\item[7] E. Celeghini, M. Rasetti and G. Vitiello,
{\it Annals of Phys.	(N.Y.)} {\bf 215} (1992) 156

\item[8] Y.N.Srivastava, G.Vitiello and A.Widom,
{\it Annals of Phys. (N.Y.)}, {\bf 238 } (1995) 200

\item[9] G.Vitiello, in {\it Quantum-like models and coherent effects }, 
R.Fedele and P.K.Shukla 
Eds., World Scintific Publ. Co., Singapore 1995, p.136 

\item[10] H. Bateman, {\it Phys. Rev.} {\bf 38} (1931), 815

\item[11] H.Umezawa, {\it Advanced field theory: micro,
macro and thermal concepts}, American Institute of Physics, N.Y. 1993\\
Y.Takahashi and H.Umezawa, { \it Collective 
Phenomena} {\bf 2} (1975) 55\\
H.Umezawa, H.Matsumoto and M.Tachiki, {\it Thermo 
	field dynamics and condensed states}, North-Holland Publ. Co., 
	Amsterdam 1982

\item[12]  H.Feshbach and Y.Tikochinsky, { \it Transact. N.Y. Acad. Sci.}
	{\bf 38} (Ser. II) (1977) 44

\item[13]  Y. Tsue, A. Kuriyama and M. Yamamura,  {\it Progr. Theor. Physics}
{\bf 91} (1994), 469

\item[14] J.M.Ziman, {\it Electrons in metals}, Taylor \& Francis Ltd., 
London 1970

\item[15] E.Abe, {\it Hopf algebras}, Cambridge Tracts in Math. {\bf 74}, 
Cambridge University Press 1980

\item[16] A.Iorio and G.Vitiello, {\it Annals of Phys. (N.Y.)}, {\bf 241 } 
(1995) 496

\item[17] S.De Martino, S.De Siena and G.Vitiello, {\it Int. J. Mod. 
Phys.} {\bf B}, Special Issue dedicated to H.Umezawa, May 1996 in print \\
E.Celeghini, S.De Martino, S.De Siena, A.Iorio, 
M.Rasetti and G.Vitiello, in preparation 

\item[18] F.C.Khanna et al. Eds., {\it Banff/CAP Workshop on thermal field 
        theory}, World Scientific Publ. Co., Singapore 1994 

\item[19]  O.Bratteli and D.W.Robinson, {\it Operator Algebras and Quantum 
    Statistical Mechanics}, Springer, Berlin 1979

\item[20] H.Umezawa and G.Vitiello, {\it Quantum Mechanics}, Bibliopolis, 
Naples 1985 \\
H.Goldstein, {\it Classical Mechanics}, Addison-Wesley Publ. Co. Inc.,
Cambridge,Mass., 1956

\item[21] C.Itzykson and J.Zuber, {\it Quantum field theory},
McGraw-Hill Book Co., N.Y. 1980

\item[22] A.Comtet, {\it Annals of Phys. (N.Y.)} {\bf 173} (1987) 185

\item[23] M.Berry and M.Robnik, {\it J. Phys. A : Math. Gen.} {\bf 19} 
(1986) 649

\item[24] V.I.Arnold and A.Avez, {\it Ergodic problems of classical mechanics},
W.A.Benjamin Inc.,N.Y. 1968

\item[25] A.A.Andronov, A.A.Witt and S.E.Haikin, {\it Vibration theory},
Moscow, Nauka, 1981

\item[26] N.L.Balazs and A.Voros, {\it Phys.Rep.}{\bf 143} (1986) 109

\item[27] I.M.Gelfand and S.V.Fomin, {\it Trans.Amer.Math.Soc.}
{\bf 2}(1955) 49

\end{itemize}

\end{document}